%% file: eprint.tex
\documentclass[12pt]{article}
\usepackage{graphicx}

\newcommand\pubnumber{NuPhys2016-Prior}
\newcommand\pubdate{\today}

\def\lip{Laborat\'orio de Instrumenta\c c\~ao e F\'isica Experimental de Particulas}

\def\Title#1{\begin{center} {\Large #1 } \end{center}}
\def\Author#1{\begin{center}{ \sc #1} \end{center}}
\def\Address#1{\begin{center}{ \it #1} \end{center}}

\newcommand\pubblock{\rightline{\begin{tabular}{l} \pubnumber\\
         \pubdate  \end{tabular}}}
\newenvironment{Abstract}{\begin{quotation}  }{\end{quotation}}
\newenvironment{Presented}{\begin{quotation} \begin{center} 
             PRESENTED AT\end{center}\bigskip 
      \begin{center}\begin{large}}{\end{large}\end{center} \end{quotation}}
\def\Acknowledgements{\bigskip  \bigskip \begin{center} \begin{large}
             \bf ACKNOWLEDGEMENTS \end{large}\end{center}}

\input econfmacros.tex

\begin{document}
\begin{titlepage}
\pubblock

\vfill
\Title{The SNO+ experiment physics goals\\
and backgrounds mitigation}
\vfill
\Author{Gersende Prior (for the SNO+ Collaboration)}
\Address{\lip}
\vfill
\begin{Abstract}
The main physics goal of the SNO+ experiment is the search for neutrinoless double-beta decay (0$\nu\beta\beta$), a rare process which if detected, will prove the Majorana nature of neutrinos and provide information on the absolute scale of the neutrino mass. Additional physics goals include the study of solar neutrinos, anti-neutrinos from nuclear reactors and the Earth's natural radioactivity as well as Supernovae neutrinos. Located in the SNOLAB (Canada) deep underground laboratory, it will re-use the SNO detector. A short phase with the detector completely filled with water has started at the beginning of 2017, before running the detector with scintillator. This paper describes in details the SNO+ sensitivity to 0$\nu\beta\beta$ decays, as well as the other physics goals. Crucial to these large volume liquid scintillator experiments, is the ability to constraint the low-energy background from radioactive decays. Methods to mitigate those are also presented.
\end{Abstract}
\vfill
\begin{Presented}
NuPhys2016, Prospects in Neutrino Physics\\
Barbican Centre, London, UK, December 12-14, 2016
\end{Presented}
\vfill
\end{titlepage}
\def\thefootnote{\fnsymbol{footnote}}
\setcounter{footnote}{0}
\section{The SNO+ experiment}
The SNO+ experiment uses the SNO $\sim$9300 photo-multipliers (PMTs). The detector was upgraded with a new hold-down ropes systems, new data-acquisition and readout systems, and modifications of its water plant. New scintillation and tellurium purification plants are also being completed at present. The detector is constituted of a geodesic steel structure of 17 m diameter which supports the PMTs. Inside, a spherical acrylic vessel of 6 m radius holds the media. The acrylic vessel is shielded by 7 kt of pure water which fills the entire cavern, additional rock shields composed of norite and granite/gabbro are surrounding the detector located 2 km underground in SNOLAB (Canada). SNO+ will have several experimental phases: a water phase, a scintillator phase with 780 tons of liquid scintillator and a scintillator loaded phase with 1.33 tons of $^{130}$Te (0.5\% $^{nat}$Te). The detector is equipped with a set of different calibration systems which are composed of deployable radioactive sources (provide an estimation of the reconstruction efficiency and the systematics uncertainties) and optical systems (measure the PMTs response and media properties such as the attenuation and scattering coefficients). See \cite{andringa,liggins,dunger,falk} for more detailed information.
\section{The SNO+ Physics}
\subsection{Neutrinoless double-beta decay with $^{130}$Te}
\begin{figure}[htb]
\centering
\begin{minipage}{1.0\textwidth}
\centering
\includegraphics[height = 6 cm]{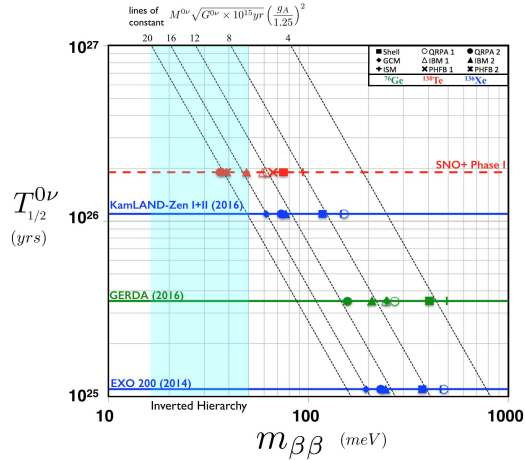}
\end{minipage}
\caption{0$\nu\beta\beta$ half-life and effective Majorana mass sensitivity for different isotopes and NME values.}
\label{fig:sensitivity}
\end{figure}

Two neutrinos double-beta decay (2$\nu\beta\beta$) is a very rare process that is permitted for 35 known natural isotopes. It was experimentally observed in 11 of them.
Neutrinoless double-beta decay (0$\nu\beta\beta$) can occur if neutrinos have a mass and they are their own anti-particles (Majorana neutrinos). The effective Majorana mass corresponds to $<m_{\beta\beta}> = |\sum m_i \times U_{ei}^{2}|$ \cite{avignone}, where m$_i$ is the mass of the neutrino eigenstate i and U$_{ei}$ the Pontecorvo-Maki-Nakagawa-Sakata (PMNS) matrix elements (i = 1,2,3). The 0$\nu\beta\beta$ decay rate can be written as follows:\\
\[\Gamma^{0\nu\beta\beta} = ln(2) \cdot (T^{0\nu\beta\beta}_{1/2})^{-1} = ln(2) \cdot G^{0\nu\beta\beta}(Q_{\beta\beta},Z) \cdot g_{A}^{4}\cdot | M^{0\nu\beta\beta}|^{2}\cdot \frac{<m_{\beta\beta}>^{2}}{m_e^2}\]
where G$^{0\nu\beta\beta}$ is the phase-space factor and M$^{0\nu\beta\beta}$ the Nuclear Matrix Elements (NME), both have values depending on the isotope chosen \cite{barea}. The limit on the isotope half-life can be described as:\\
\[T^{0\nu\beta\beta}_{1/2} = \frac{N\cdot ln(2)}{n_{\sigma}}\cdot \frac{f(\delta_{\epsilon})\cdot t}{\sqrt{(b\cdot M + c)\cdot \delta E\cdot t}}\]
with N corresponding to the total number of isotope nuclei, n$_\sigma$ the number of standard deviation, f($\delta_{\epsilon}$) the energy window acceptance fraction, t the time, M the isotope mass in kg, $\delta$E the energy window in keV, b the background counts in (keV.kg.yr)$^{-1}$ and c the backgrounds counts in (keV.yr)$^{-1}$. Background b (e.g. U/Th) scales with the isotope quantity whereas c is independent of the isotope (e.g. solar $^8$B). $^{130}$Te has been chosen by the SNO+ collaboration for the following properties: its high natural abundance (34.08\%), its 2$\nu\beta\beta$ long half-life (7.0 $\times$ 10$^{20}$ yr) and the absence of inherent absorption lines. However, its energy end-point Q$_{\beta\beta}$ = 2.527 MeV, which requires a very detailed knowledge of the backgrounds falling in the Region of Interest (ROI).\\
The 0$\nu\beta\beta$ half-life as a function of the effective Majorana neutrino mass for different isotopes is given in Figure \ref{fig:sensitivity}.
\subsection{Backgrounds mitigation}
\begin{figure}[htb]
\centering
\begin{minipage}{0.5\textwidth}
\centering
\includegraphics[height = 4 cm]{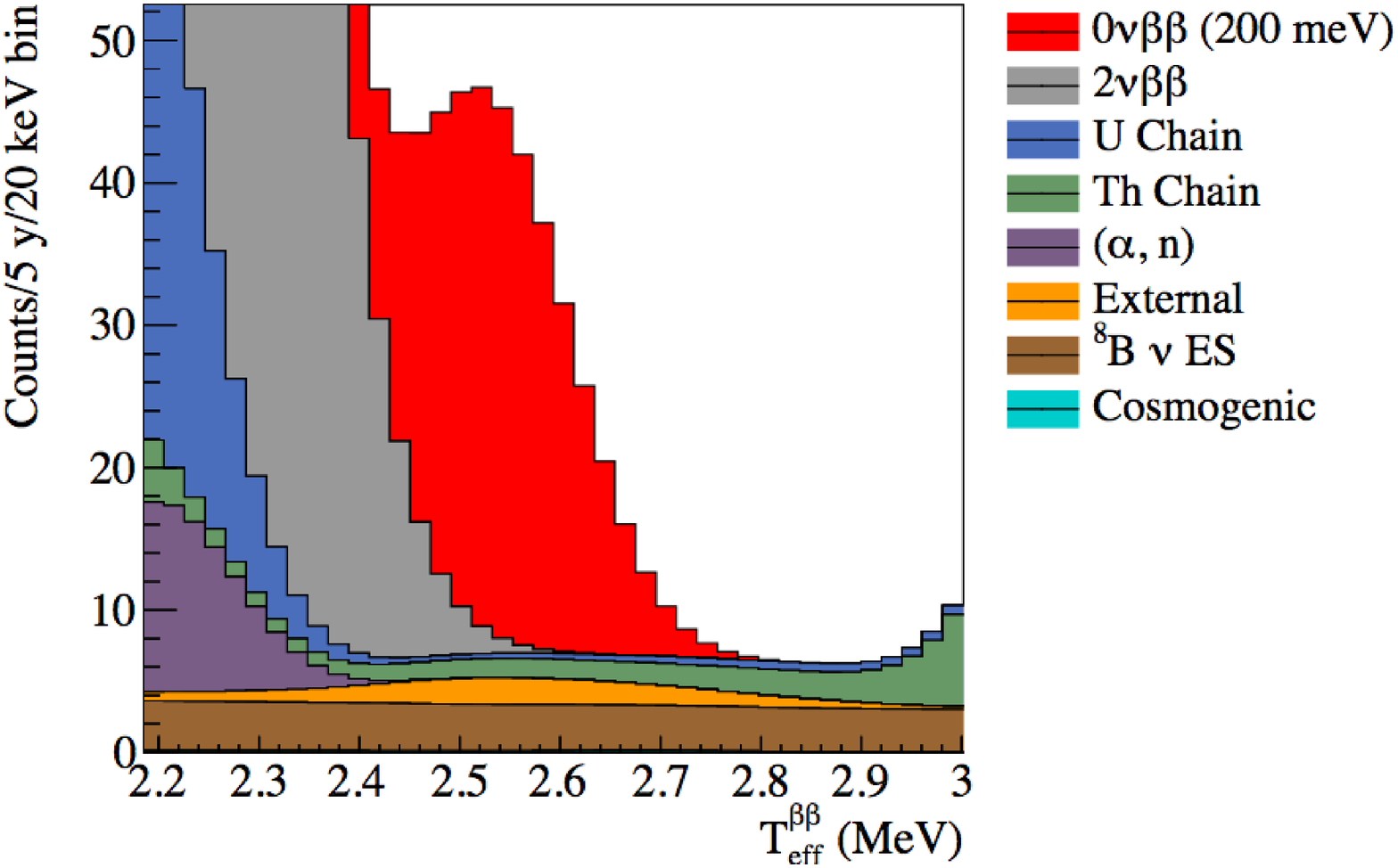}
\end{minipage}%
\begin{minipage}{0.5\textwidth}
\centering
\includegraphics[height = 4 cm]{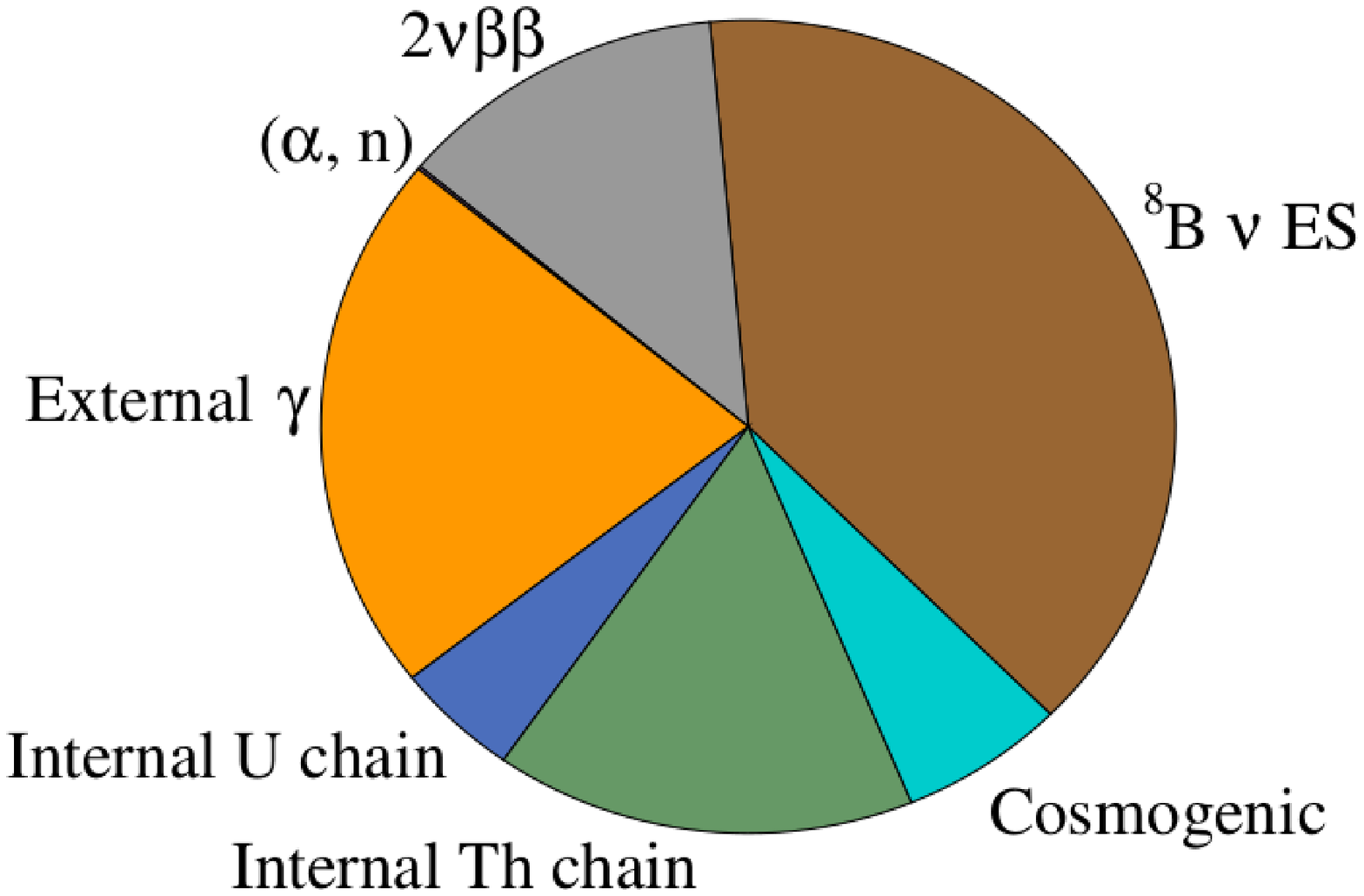}
\end{minipage}%
\caption{Stacked signal+backgrounds spectrum for 5 years data with 0.5\% $^{nat}$Te and a FV cut R = 3.5 m (left), background budget in the first year of data (right).}
\label{fig:signal_background}
\end{figure}

A novel technique for loading the scintillator cocktail (LAB+PPO) uses Te-Diol, which will provide a light yield of 390 hits/MeV. The spectrum of an hypothetic 0$\nu\beta\beta$ signal for a effective Majorana mass of 200 meV and backgrounds for 5 years of data-taking with 0.5\% $^{nat}$Te loading and a FV cut R = 3.5 m is given in Figure \ref{fig:signal_background} (left).
The expected sensitivies for the 0$\nu\beta\beta$ half-life T$^{0\nu\beta\beta}_{1/2}$ is $>0.8\times 10^{26}$ yr (1.96$\times 10^{26}$ yr) for 1 year (5 years) data-taking\footnote{The values G$^{0\nu\beta\beta}$= 3.69$\times 10^{-14}$ yr$^{-1}$, g$_{A}$ = 1.269 and M$^{0\nu\beta\beta}$ = 4.03 (IBM-2) have been used.}. The background budget can be found in Figure \ref{fig:signal_background} (right) representing 13.4 counts/yr in FV and ROI. The 2$\nu\beta\beta$ background is irreducible but will be constrained by using an asymetric, [$\mu-0.5\sigma$,$\mu+1.5\sigma$], ROI, however the energy resolution is limited. Background from ($\alpha$,n) will be rejected using coincidence-based cuts with an expected efficiency $>$ 99.6\% (90\%) for the prompt (delayed) signal. External $\gamma$ backgrounds will be identified using FV and PMTs time distribution cuts. For the U and Th chain, the dominant background from $^{214}$BiPo and $^{212}$BiPo decays will be rejected using coincidence-based cuts with 100\% rejection power for events in separate triggers. Purification techniques and long term underground storage will help eliminate backgrounds from cosmogenics. Finally the continuous background from elastically scattered electron from the solar $^{8}$B interaction will be normalized using published data.
\subsection{Other physics}
\begin{figure}[htb]
\centering
\begin{minipage}{0.5\textwidth}
\centering
\includegraphics[height = 3.5 cm]{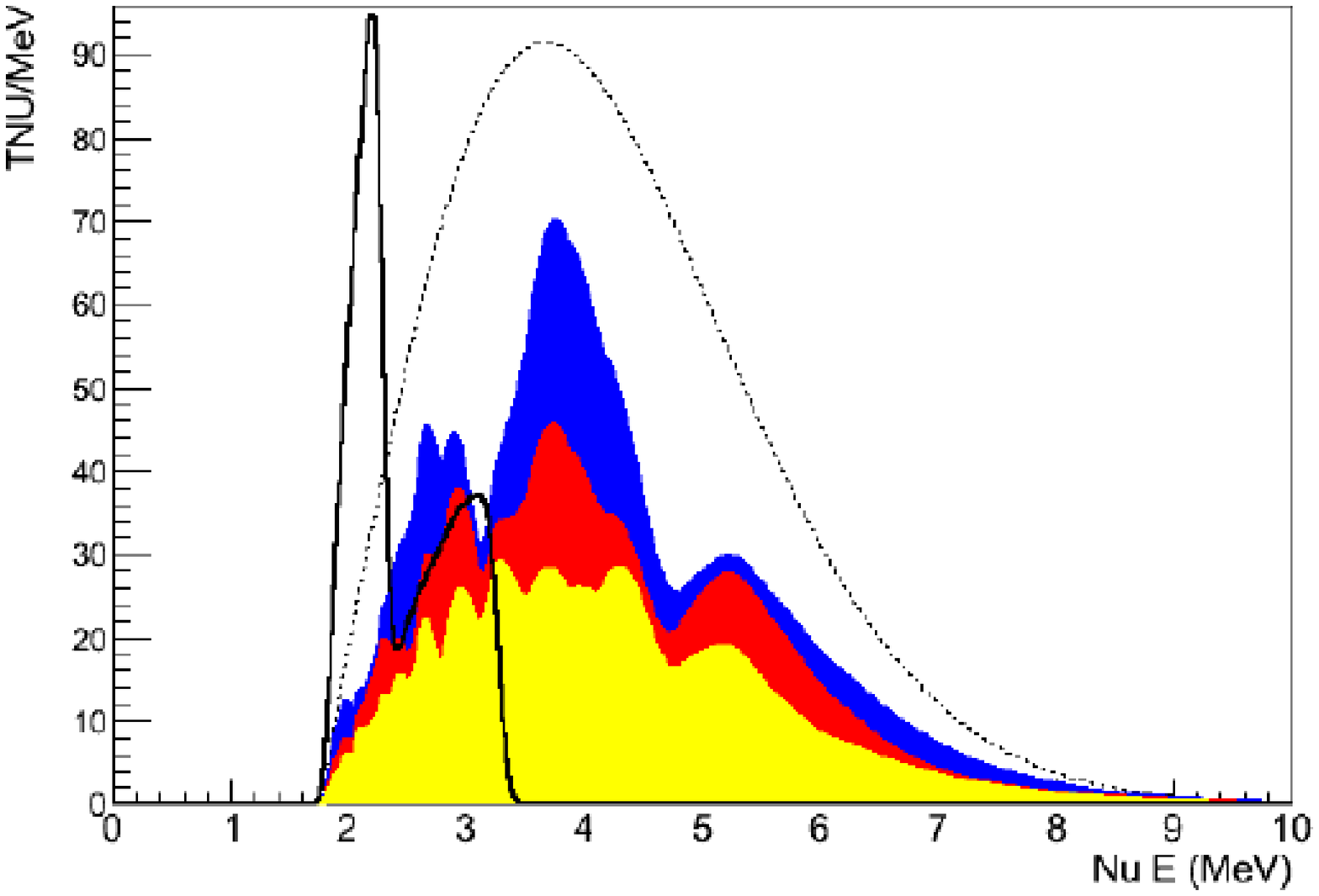}
\end{minipage}%
\begin{minipage}{0.5\textwidth}
\centering
\includegraphics[height = 3.5 cm]{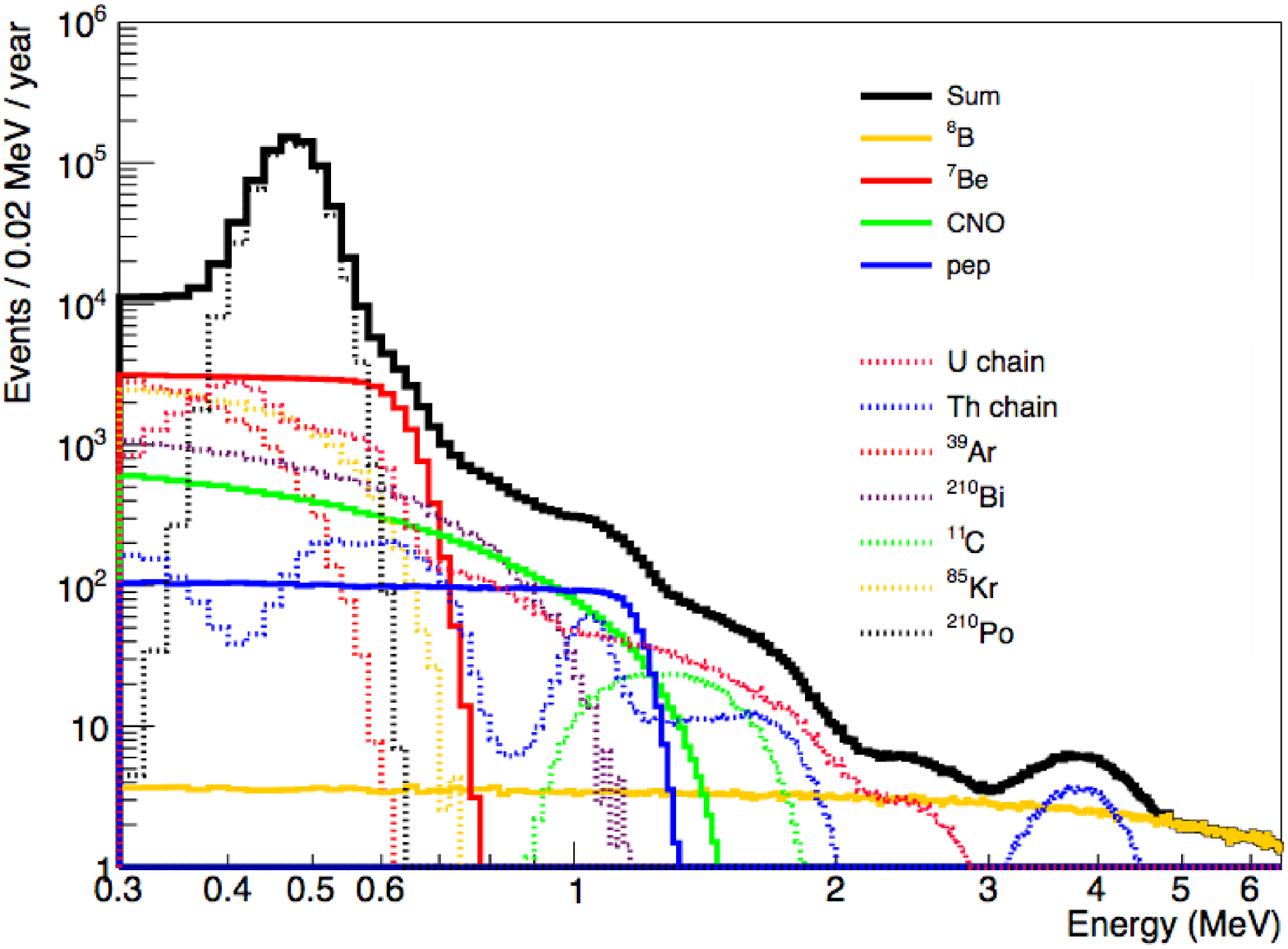}
\end{minipage}%
\caption{Visible anti-neutrino energy spectrum for 10$^{52}$ proton-year \cite{andringa} (left) and solar neutrino spectrum (right).}
\label{fig:antinu_solar}
\end{figure}

\begin{figure}[htb]
\centering
\begin{minipage}{0.5\textwidth}
\centering
\includegraphics[height = 3.5 cm]{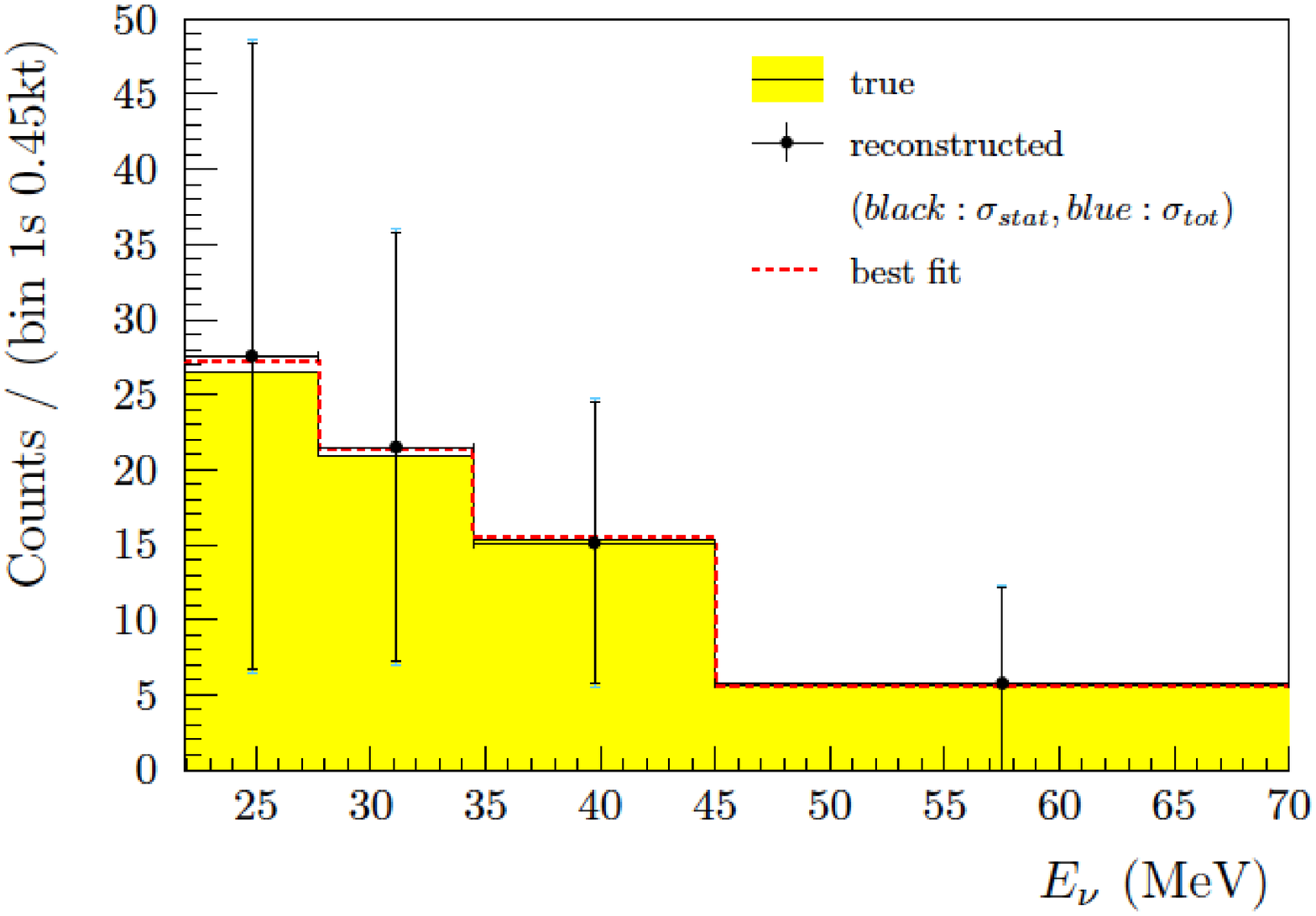}
\end{minipage}%
\begin{minipage}{0.5\textwidth}
\centering
\includegraphics[height = 3.5 cm]{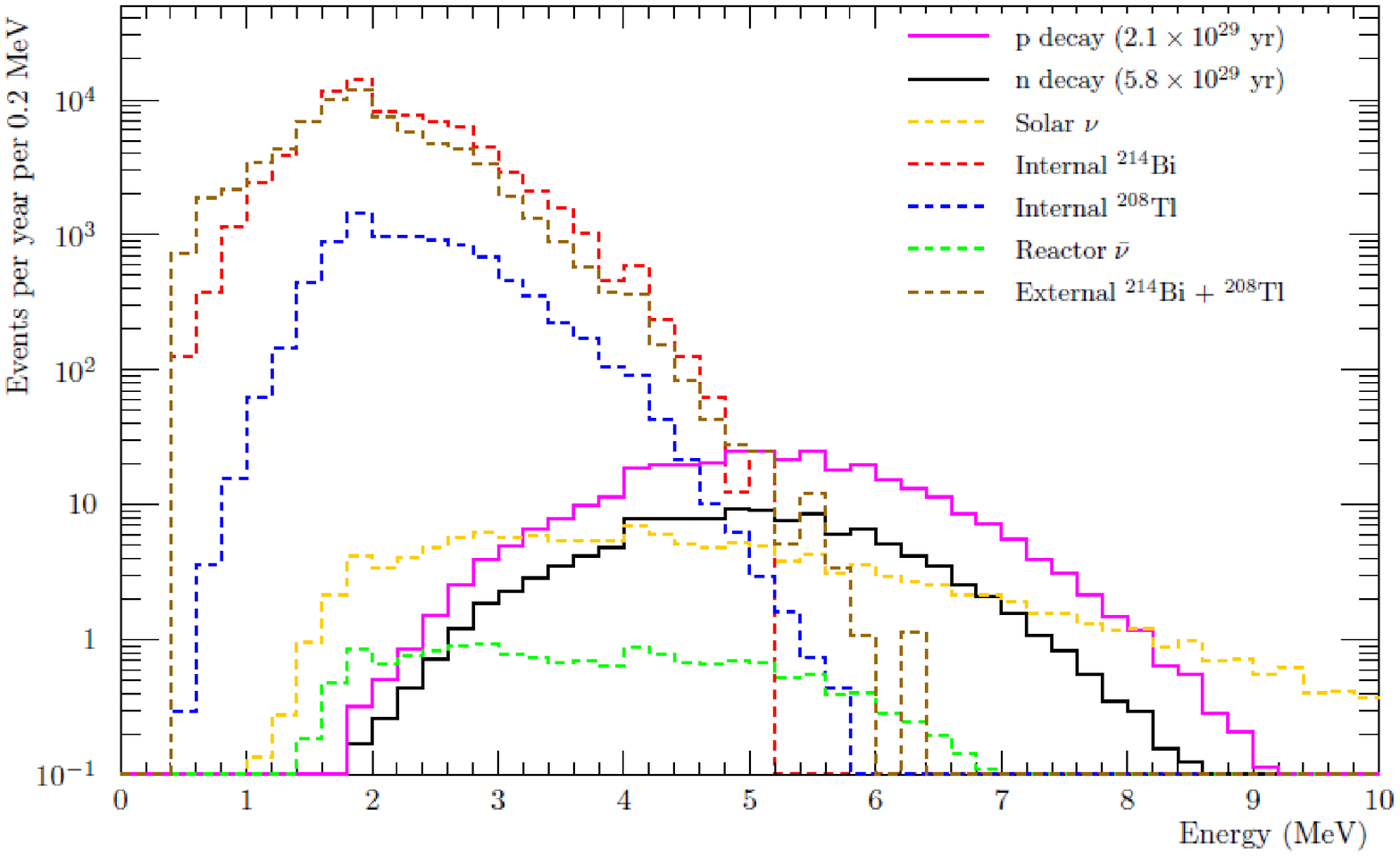}
\end{minipage}%
\caption{Neutrino energy spectra from a Supernova explosion \cite{krosigk} in the $\nu$-p ES channel (left) and nucleon decay spectra with a FV cut R = 5.5 m (right).}
\label{fig:sn_ndecay}
\end{figure}

In addition to the search for 0$\nu\beta\beta$ decays, SNO+ has several other physics goals described below. Anti-neutrinos coming from nearby nuclear reactors will be measured (see Figure \ref{fig:antinu_solar}, left) in the pure and Te-loaded liquid scintillator phases. The neutron delayed capture (2.2 MeV $\gamma$) from the inverse-beta-decay (IBD) reaction is used to tag such events. The expected precision to the neutrino oscillation parameter $\Delta m^{2}_{12}$ is 0.2$\times$10$^{-5}$ eV$^2$ for 7 years of data-taking, similar to the precision obtained by KamLAND \cite{gando}.
Detection of Solar neutrinos is done by recording the recoil electron signal from the neutrino elastic (ES) scattering in the pure scintillator phase (see Figure \ref{fig:antinu_solar}, right). The expected precision on the different fluxes is of 7.1\% for $^{8}B$, 3.3\% for $^{7}$Be, 8.9\% for pep and 15\% for CNO for 1 year data. Neutrinos from Supernovae located in the vicinity of 10 kpc (see Figure \ref{fig:sn_ndecay}, left) can be detected in all phases of the SNO+ experiment.
A search for invisible modes such as the neutron decay from $^{16}$O into 3$\nu$ with emission of several $\gamma$s \cite{ejiri} will be performed in the water phase (see Figure \ref{fig:sn_ndecay}, right). The expected limit on the neutron and proton lifetimes for 6 months of data is $\tau_{n}\ge1.25\times10^{30}$ yr and $\tau_{p}\ge1.38\times10^{30}$ yr.
\Acknowledgements
This work is funded by Funda\c c\~ao para a Ci\^encia e a Tecnologia (FCT-Portugal) within the project grant IF/00863/2013/CP1172/CT0006 and by the Portuguese State.
\end{document}

%% file: econfmacros.tex
\def\beq{\begin{equation}}
\def\eeq#1{\label{#1}\end{equation}}
\def\eeqn{\end{equation}}

\def\beqa{\begin{eqnarray}}
\def\eeqa#1{\label{#1}\end{eqnarray}}
\def\eeqan{\end{eqnarray}}

\let\bar=\overbar

\def\Dslash{\not{\hbox{\kern-4pt $D$}}}
\def\dslash{\not{\hbox{\kern-2pt $\del$}}}

\def\msb{{\bar{\ssstyle M \kern -1pt S}}}

